\documentclass{moriond}

\def\be{\begin{equation}}
\def\ee{\end{equation}}
\def\bea{\begin{eqnarray}}
\def\eea{\end{eqnarray}}

\newcommand{\Photo}{}

\begin{document}
\vspace*{4cm}
\title{CONSTRAINING GENERAL MODIFICATIONS OF GRAVITY \newline DURING REIONISATION}

\author{ C. HENEKA$^{1,2}$ and L. AMENDOLA$^2$ }

\address{$^1$ Scuola Normale Superiore, Piazza dei Cavalieri 7, 56126 Pisa, Italy \\
$^2$ Institut f\"ur Theoretische Physik, Ruprecht-Karls-Universit\"at Heidelberg, \\ Philosophenweg
16, 69120 Heidelberg, Germany}

\maketitle\abstracts{
Intensity mapping enables us to test cosmology and fundamental physics during the Epoch of Reionisation, with power spectra of line fluctuations probing a large range of scales and redshifts. Cosmological volumes of 21cm line fluctuations in general modified gravity scenarios are presented, where additional parameters are the initial condition parameter $\alpha$ of matter perturbations and the scale-dependent modified gravity parameter $Y$ that measures deviations from General Relativity in the Poisson equation. For upcoming surveys like the SKA we forecast, using the power spectra derived from our simulations, the ability of intensity mapping to constrain modifications of gravity during reionisation, as well as investigate correlations with astrophysical parameters.}

\section{Introduction}
Intensity mapping, where fluctuations of line emission are mapped out, has the potential to constrain both astrophysics and cosmology over a large range of scales and redshifts. It posits the unique opportunity to push measurements to high redshifts of the Epoch of Reionisation (EoR) where our cosmology and model of gravity is largely unconstrained so far. During this EoR the first stars and galaxies ionise again the before neutral medium around them, reaching a fully ionised Universe at a redshift of $z\sim6$.~\cite{2015MNRAS.447..499M} 

Our goal in this proceeding is to demonstrate our ability to constrain general modifications of gravity at redshifts of reionisation in order to better understand cosmic acceleration. We forecast constraints attainable for experiments like the upcoming Square Kilometre Array (SKA) that aims to detect 21cm fluctuations tracing neutral hydrogen.~\cite{2015aska.confE...1K}
In order to stay as model-independent as possible, we derive constraints on the modification to the standard Poisson equation $Y\left( a,k\right)$, with $Y=1$ in General Relativity (GR), together with the initial condition parameter $\alpha$, with $\alpha=1$ for matter domination at early times.

\subsection{Linear growth for general modifications of gravity}\label{sec:growth}
For the background we evolve the Hubble parameter $E\left(a\right)$ with scale $a$, where we have chosen a CPL parametrisation for the time evolution of $w$. Fiducial model parameters are $\sigma_8=0.815$, $h=0.678$, $\Omega_\mathrm{r}=8.6\times10^{-5}$, $\Omega_\mathrm{m,0}=0.308$, $w_\mathrm{0}=-1$, $w_\mathrm{a}=0$. General modifications of gravity enter at the linear regime as the effective gravitational strength $Y$ and the initial condition parameter $\alpha$. We evolve the growth of matter perturbations in this general scenario as~\cite{Amendola:2012ky}
\begin{equation}
\delta''_\mathrm{m} + \left( 2+\frac{E'}{E}\right)\delta'_\mathrm{m} = \frac{3}{2}\frac{\delta_\mathrm{m}}{a^3E^2}\Omega_\mathrm{m,0}Y ,
\label{eq:pert2}
\end{equation}
with $\alpha=\delta'_{in}/\delta_{in}$ at early time  (prime denotes here derivatives after $\log a$).  
In general, both $Y$ and $\alpha$ can be scale- and time-dependent functions. They are both equal to one in GR with $Y=1$ and $\alpha=1$. Here we treat $\alpha$ as a constant, as well as constrain a $Y$ that is constant during the EoR in this first study. The growth function defined as $G\left( a,k\right)=\delta_m\left( a,k\right)/\delta_m\left(1,k\right)$ is calculated via Eq.~(\ref{eq:pert2}) and normalised to the same growth at the time of recombination, or alternatively the same $\sigma_8$ as constrained by the Cosmic Microwave Background (CMB).
The growth evolution of matter perturbations is then inputted in our semi-numerical code to calculate the 21cm signal as described in the following section.

\subsection{Simulations of the 21cm signal in general modified gravity}
We simulate maps of 21cm line emission with the semi-numerical code 21cmFAST,\footnote{https://github.com/andreimesinger/21cmFAST} which creates density, velocity, as well as ionisation fields. We modified this code as described in the previous section to incorporate growth as evolved in our modified gravity scenario. 

The 21cm brightness temperature offset $\delta T_\mathrm{b}$ between spin gas temperature and CMB background temperature at redshift $z$ and position $\bf{x}$ is obtained via
\begin{equation}
\delta T_\mathrm{b} \left({\bf x}, z \right)  \approx  27x_\mathrm{HI}\left( 1+\delta_\mathrm{nl}\right)\left( \frac{H \left(z\right)}{\mathrm{d}v_\mathrm{r}/\mathrm{d}r + H\left(z\right)}\right) \left( \frac{1+z}{10} \frac{0.15}{\Omega_\mathrm{m,0} h^2}\right) \left( \frac{\Omega_\mathrm{b} h^2}{0.023}\right) \mathrm{mK} , \label{eq:Tb}
\end{equation}
with neutral fraction $x_\mathrm{HI}$, density contrast $\delta_\mathrm{nl}=\rho / \bar{\rho}_0 -1$, Hubble function $H\left( z\right)$, comoving gradient of line of sight velocity $\mathrm{d}v_\mathrm{r}/\mathrm{d}r$, as well as present-day matter density $\Omega_\mathrm{m,0}$, present-day baryonic density $\Omega_\mathrm{b}$, and Hubble factor $h$. This relation Eq.~(\ref{eq:Tb}) for the 21cm brightness temperature is valid in the post-heating limit $T_{\gamma} \ll T_\mathrm{S}$.
Fiducial reionisation model parameters are chosen as mean free path of ionising radiation $R_\mathrm{mfp} =20\,$Mpc, typical virial temperature $T_\mathrm{vir}=3\times10^4\,$K, and ionising efficiency $\zeta=20$. We also choose the relevant parameters for the heating history, the efficiency of X-ray heating $\zeta_\mathrm{x}$ and the mean baryon fraction in stars $f_{*}$, as $\zeta_\mathrm{x} = 2\times10^{56}$ and $f_{*} =0.05$, in accordance with current bounds. 

\begin{figure}
\begin{center}
\includegraphics[width=0.32\columnwidth]{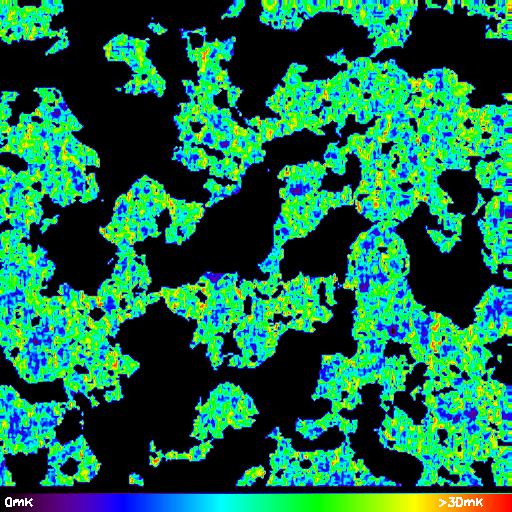}
\includegraphics[width=0.32\columnwidth]{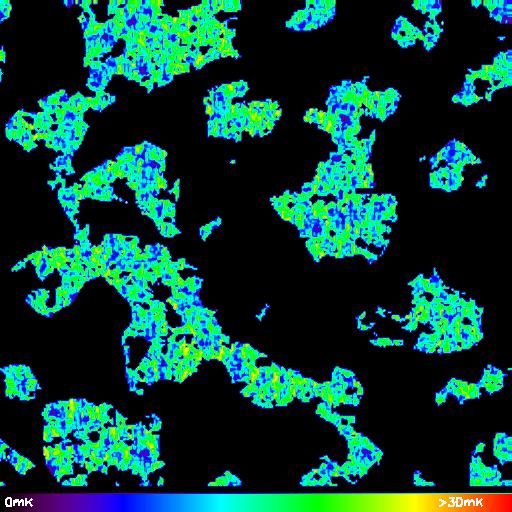}
\includegraphics[width=0.32\columnwidth]{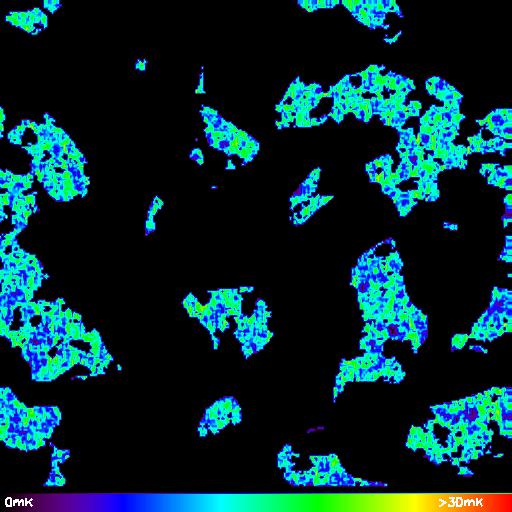}
\caption{Examples of 300$\,$Mpc simulation boxes of 21cm emission at redshift $z=7$, for our fiducial cosmology with $\alpha=1$, but varying $Y=0.99$ (left), $Y=1.00$ (middle) and $Y=1.01$ (right).
}
\label{FIG:Box21_Y_CMB}
\end{center}
\end{figure}

Temperature fluctuations $\delta_{21} \left({\bf x},z\right)$ on the simulated grid at position $\bf{x}$ and redshift slice $z$ are then calculated as $\delta_{21} \left({\bf x},z\right) = \delta T_\mathrm{b} \left( {\bf x},z\right) /  \bar{T}_{21}\left( z\right) - 1$,
with average 21cm brightness temperature $\bar{T}_{21}\left( z\right)=<\delta T_\mathrm{b}>_{{\bf x}}$. In figure~\ref{FIG:Box21_Y_CMB} simulated boxes of 300$\,$Mpc box size for fluctuations in 21cm brightness temperature are depicted at redshift $z=7$. We can see going from left ($Y=0.99$) to right ($Y=1.01$), that with higher effective gravitational strength $Y$ reionisation has progressed more at the same redshift, with ionised (dark) patches becoming more prominent.

\section{Constraints on general modifications of gravity}

\begin{figure}
\begin{center}
\includegraphics[width=0.38\columnwidth]{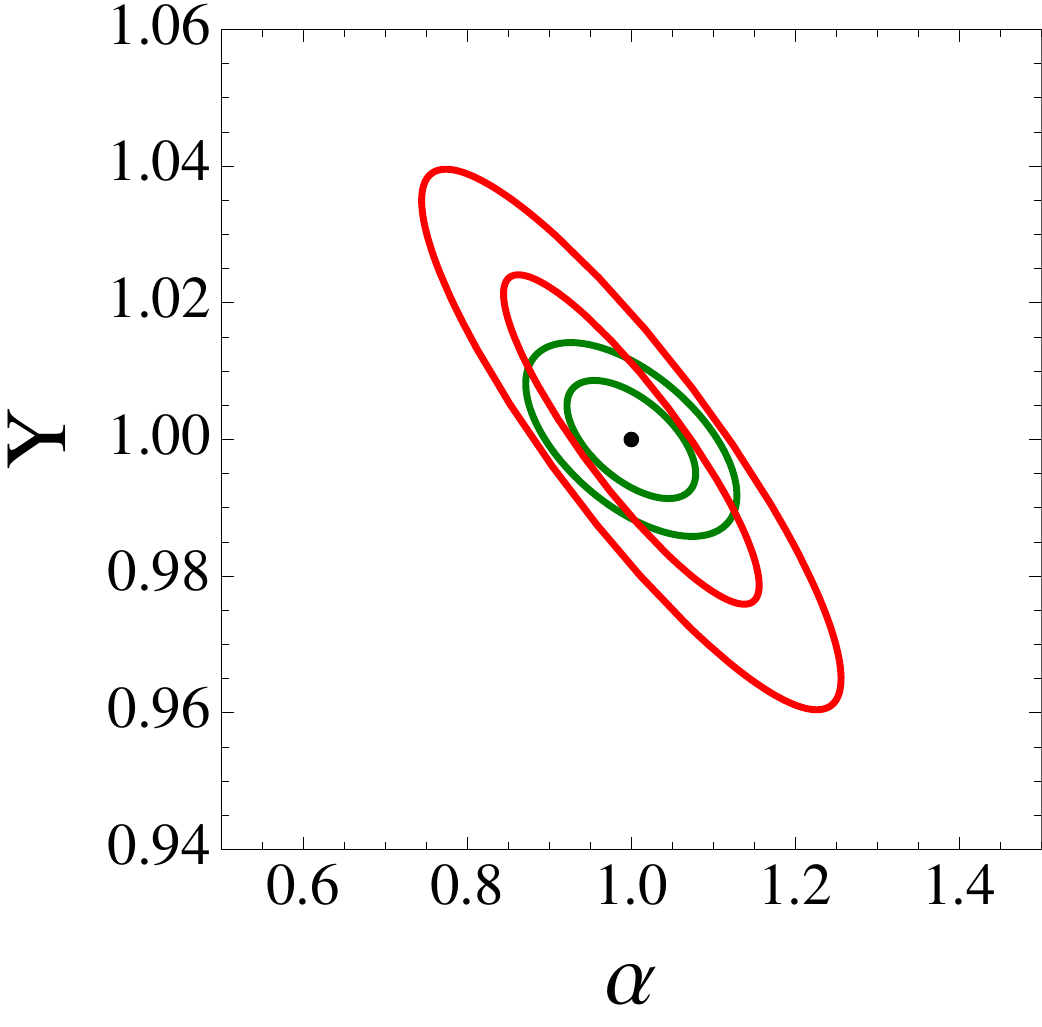}
\hspace{0.5cm}
\includegraphics[width=0.51\columnwidth]{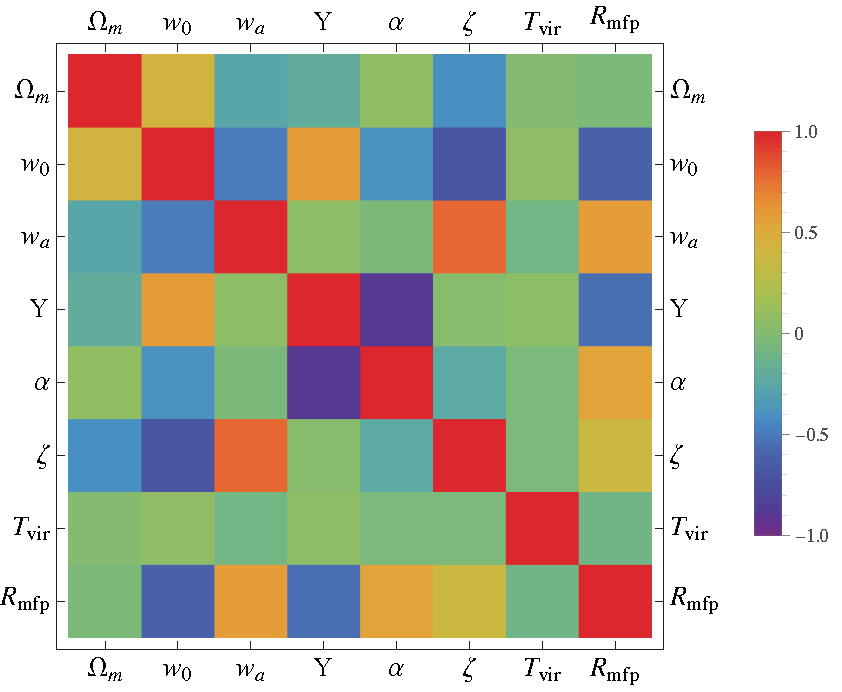}
\caption{ Left: Two-dimensional 1- and 2-$\sigma$ confidence contours for modified gravity parameters $Y$ and $\alpha$.  Green contours assume the 'idealised' scenario of no foregrounds for cosmological parameters-only, red contours have reionisation model parameters and foreground removal added in the Fisher forecast. Right: Correlation matrix from our Fisher analysis, for the combined cosmological and reionisation set of model parameters, with the foreground wedge removed. For more details please see section~\ref{sec:Fisher} and~\ref{sec:corr}.}
\label{FIG:corrR}
\end{center}
\end{figure}

\subsection{Parameter constraints}\label{sec:Fisher}
We perform a Fisher matrix forecast for the set of cosmological parameters $\left(\Omega_\mathrm{m,0}, w_0, w_\mathrm{a}, Y, \alpha \right)$, as well as with the reionisation parameters $\left( \zeta, T_\mathrm{vir}, R_\mathrm{mfp} \right)$ added to the parameter vector. Parameters are assumed to be constant and scale-independent during the EoR. For each parameter set and each redshift, the corresponding 21cm emission was simulated as described in the previous section, then 21cm power spectra are extracted. We combine constraints for measurements of the 21cm power spectrum in 6 redshift bins from $z=6$ to $z=11$ in steps of $\Delta z=1.0$. For our error estimate we assume a SKA stage 1 like intensity mapping survey to measure the 21cm power spectrum. Survey size, cosmic variance, thermal noise and instrumental resolution are accounted for. We also explore the case of removing the 21cm foreground wedge. 

The left panel of figure~\ref{FIG:corrR} shows the corresponding confidence contours derived for the parameters $\left( Y, \alpha \right)$, without the removal of the 21 cm foreground wedge in the case of cosmological parameters only (green), as well with reionisation parameters and foreground removal included in the Fisher analysis (red). We can see that foreground removal and inclusion of reionisation parameters in a more realistic scenario decreases the precision by about a factor of two, from $\Delta Y=0.006$ and $\Delta \alpha = 0.05$ to $\Delta Y = 0.016$ and $\Delta \alpha = 0.10$. In general, constraints at the order of percent for $Y$ and tens of percent for $\alpha$ are attainable. We note, that imposing more conservative high-k cuts for the non-linear regime can degrade constraints, and improving the modelling of non-linear scales then proves crucial.

\subsection{Correlations between model parameters}\label{sec:corr}
To check how strongly fixing reionisation model parameters impacts the Fisher forecast of our cosmology, besides affecting 1$\sigma$ marginalised errors as shown in the previous section, we calculated the covariance matrix ${\bf C}$ as the inverse of the Fisher matrix for the full parameter vector $p=\left( \Omega_\mathrm{m,0}, w_0, w_\mathrm{a}, Y, \alpha, \zeta, T_\mathrm{vir}, R_\mathrm{mfp}\right)$. The correlation matrix $P_{ij} = C_{ij} /\sqrt{C_{ii} C_{jj}}$  is 1 for perfect correlation and -1 for perfect anti-correlation between two elements of our parameter vector p. 

In figure~\ref{FIG:corrR} (right panel)\cite{Heneka:2018ins} we show the correlation matrix for the full parameter vector.
As compared to the cosmology-only case correlations between cosmological parameters are only slightly affected by adding reionisation model parameters. As one example for correlations between our modified gravity parameters and reionisation parameters, $Y$ that regulates the strength of gravity is anti-correlated with the mean free path $R_\mathrm{mfp}$, with stronger gravity meaning a shorter mean free path, but shows no significant correlation with typical halo virial temperature and ionising efficiency. We also note that removing the foreground wedge or imposing other high-k cuts helps to de-correlate parameters.

\section{Conclusions and Outlook}
Here we demonstrated the ability of upcoming intensity mapping experiments like the SKA that probe the EoR to put unique constraints on the deviation $Y$ from the standard Poisson equation and the initial condition parameter $\alpha$. We did so by including the  growth history for general modifications of gravity into semi-numerical simulations of 21cm line emission. 

As shown via Fisher matrix forecast, during the EoR constraints on $Y$ can reach the percent level, as well as the tens of percent level for $\alpha$. We note that whether this precision can be reached depends on foreground treatment and treatment of the (mildly) non-linear regime. Adding reionisation parameters slightly degrades constraints, but this effect is sub-dominant as soon as adding tomographic bins in redshift lifts degeneracies. Also measurements of reionisation parameters, for example by including information from other lines than 21cm like Lyman-alpha and their cross-correlations, have the prospect to improve constraints on cosmology and modified gravity, opening the door to measure time-or scale-dependencies of modifications even wider.

\section*{Acknowledgments}
We thank the DFG for supporting this work through the SFB-Transregio TR33 ``The Dark Universe".

\section*{References}

\bibliography{ref_growth.bib}

%
%
%
%

\end{document}